\documentclass[conference]{IEEEtran}
\IEEEoverridecommandlockouts
\usepackage{cite}
\usepackage{amsmath,amssymb,amsfonts}
\usepackage{algorithmic}
\usepackage{graphicx}
\usepackage{textcomp}
\usepackage{xcolor}
\def\BibTeX{{\rm B\kern-.05em{\sc i\kern-.025em b}\kern-.08em
		T\kern-.1667em\lower.7ex\hbox{E}\kern-.125emX}}

\makeatletter
\def\ps@IEEEtitlepagestyle{
	\def\@oddfoot{\mycopyrightnotice}
	\def\@evenfoot{}
}
\def\mycopyrightnotice{
	{\footnotesize
		\begin{minipage}{\textwidth}
			\centering
			Copyright~\copyright~ ICCCAS2018. Personal use of this material is permitted. Permission from IEEE/VDE must be obtained for all other uses, in any current or future media, including reprinting/republishing this material for advertising or promotional purposes, creating new collective works, for resale or redistribution to servers or lists, or reuse of any copyrighted component of this work in other works.
		\end{minipage}
	}
}

\begin{document}

\title{Performances of C-V2X Communication on Highway under Varying Channel Propagation Models}

\author{\IEEEauthorblockN{Donglin Wang}
	\IEEEauthorblockA{\textit{University of Kaiserslautern} \\
		Kaiserslautern, Germany \\
		dwang@eit.uni-kl.de}
	\and
	\IEEEauthorblockN{Raja R.Sattiraju}
	\IEEEauthorblockA{\textit{University of Kaiserslautern} \\
		Kaiserslautern, Germany \\
		sattiraju@eit.uni-kl.de}
	\and
	\IEEEauthorblockN{Hans D.Schotten}
	\IEEEauthorblockA{\textit{University of Kaiserslautern} \\ 
			          \textit{German Research Center} \\ 
			          \textit{for Artificial Intelligence} \\
	    Kaiserslautern, Germany \\
		Hans\_Dieter.Schotten@dfki.de}
}

\maketitle

	\begin{abstract}
		In recent decades, both the industry and the academy society are sparing no efforts to develop and standardize the Cellular-Vehicle-to-Everything (C-V2X) communication which is one of the prominent emerging services of the next generation of wireless network (5G). C-V2X communication is used for information exchange among the traffic participants with network-assisted which can reduce traffic accidents and improve traffic efficiency. And it is also the primary enabler for cooperative driving. But C-V2X communication has to meet different Quality of Service (QoS) requirements (e.g., ultra-high reliability (99.999\%) and ultra-low latency). 
	\end{abstract}
	
	\begin{IEEEkeywords}
		C-V2X, 5G cellular network, channel propagation models
	\end{IEEEkeywords}
	
	\section{Introduction}
	In modern society, the rapid development of transportation system has led to certain issues such as the traffic congestion, fuel consumption, pollution, and long travel time [1]. In order to address these problems, the Cooperative Intelligent Traffic System (C-ITS) can facilitate the cooperative driving applications, such as platooning and highly-automated driving by warning the driver of dangerous situations and intervening through automatic braking or steering to help the drivers avoid the traffic accidents [2]. So the C-ITS is expected to significantly reduce the travel time, fuel consumption, and CO2 emissions while increasing road safety and traffic efficiency. The C-ITS system relies on the timely and reliable exchange of information among traffic participants, (e.g., vehicles, Road-Side Units (RSUs)), pedestrians and the network [3]. In C-ITS, C-V2X communication is a key element which enables data exchange to make traffic safer and more efficient. In order to support C-V2X communication, 5G should be able to provide a solution to support the high-reliability and high-availability, in the field of latency, Packet Reception Ratio (PRR) and other QoS parameters [4]. A lot of researches proposed direct C-V2X communication [3][4][5], which means data packets are transmitted directly from the Transmitter (Tx) to the Receivers (Rxs) without going through the network infrastructures. Since no network infrastructures involved there is lower system transmission delay thereby leading to a better system performance. For example, the IEEE 802.11p protocol has been proposed by the European Telecommunication Standards Institute (ETSI) Intelligent Transport Systems (ITS), which acts as the air interface for the direct C-V2X communication. But since there is no central entity (e.g., Base Station (BS)), packets transmission collision problem cannot be avoided [5]. Moreover, in 3GPP, PC5 has been proposed to facilitate the direct C-V2X communication which is usually referred to as the sidelink [6]. Up to 3GPP release 14, the sidelink communication is connection-less which means there is no Radio Resource Control (RRC) connection over PC5 air interface. However, in [7], C-V2X communication transmitting data packets through the cellular network infrastructures has been considered based on the legacy 4th Generation cellular network (i.e., LTE network). LTE-Uu interference facilitates the C-V2X communication and a higher end-to-end (E2E) latency compared to the direct C-V2X communication can be foreseen [7]. The performances of the direct C-V2X communication under varying channel propagation models are provided. Different radio propagation models could have varying effects on the system performance.
	 \begin{figure*}[htbp]
	 	\centering
	 	\includegraphics[width=\linewidth]{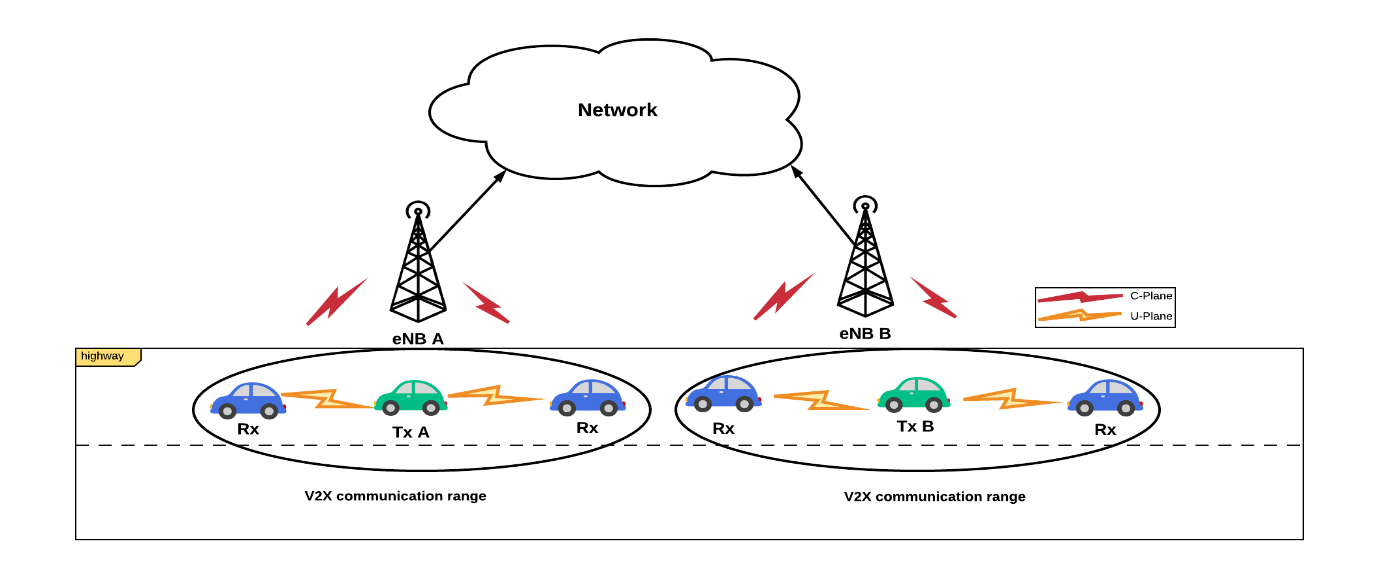}
	 	\caption{Direct C-V2X Communication with Network Assistance on a Highway}
	 	\label{fig}
	 \end{figure*}
	
	\section{System Description}
	The direct C-V2X communication through sidelink is a mode of communication whereby a User Entity (UE) can directly communicate with other UEs in its proximity over the PC5 air interface proposed in 3GPP [5]. This communication is a point-to-multipoint communication where several Rxs try to receive the same data packets transmitted from the Tx.  As shown in Fig.1, network-assisted direct C-V2X transmission model is implemented in this work under a highway scenario for data packets transmission. And all UEs are connected to BSs and the UE radio architecture consisted of the U-Plane and C-Plane is provided for the C-V2X communication. In this work, the Tx directly transmits its data packets to the surrounding Rxs in the communication range of the Tx in the U-Plane. Therefore, without involving the C-Plane, the direct C-V2X communication is efficient from the latency aspect. And all UEs are connected to the operator network in the C-plane where can provide network control for the direct C-V2X communication.  Also, in order to enable the fully automated driving and avoid accidents on highway, a C-V2X communi-cation range on a highway scenario is required to be up to 1000 meters [8]. And this communication mode is supported no matter whether the UE is under the coverage of the cellular network or not. In addition, a prerequisite is that the UE must be authorized and allowed to use the Proximity Services (ProSe) which is the logical function used for network related actions required for ProSe [9].  In 3GPP, there are two sidelink transmission modes to assign radio resources to C-V2X Txs. In the sidelink transmission mode 3, the transmission resource is scheduled by the cellular network and therefore network can allocate the same radio resource for different Txs for their direct C-V2X communication. Sidelink transmission mode 3 is only available when the vehicles are under cellular coverage. To assist the resource allocation procedure at the BS, UE context information (e.g., traffic pattern geometrical information) can be reported to BS. In the sidelink transmission mode 4, a Tx in C-V2X communication can autonomously select a radio resource from a resource pool which is either configured by network or pre-configured in the user device for its direct C-V2X communication over PC5 interface. In contrast to mode 3, transmission mode 4 can operate without cellular coverage. In this work, only transmission mode 3 is utilized for the direct C-V2X communication through sidelink which means all UEs are under the coverage of the cellular network and network control the resource allocation.
	
	\section{System-level Simulations}
	
	In order to evaluate the proposed direct C-V2X communication over sidelink by applying different channel models, a system-level simulator has been implemented to inspect on the performance of the direct C-V2X communication on a highway. In this section, we provide the detailed simulation assumptions for the direct C-V2X communication. Other system parameters and guideline are provided in [10].
	
	\subsection{Environment Model}\label{AA}
	
	In this work, a highway scenario is considered to analyze the system performance. And we also consider a more special test environment based on the highway which assumes that all vehicles on the highway with the same speed and the same Inter-Vehicle-Distance (IVD). The main configuration parameters are including a single-directional highway which is 3-lanes with 20 kilometers [3]. 
	
	\subsection{deployment Model}
	
	BSs are deployed with an Inter-Site-Distance (ISD) of 6 kilometers alongside the highway to provide C-Plane connections to the UEs of the direct C-V2X communication. And we assume the IVD is 10 meters or 15 meters to inspect the system performance with different overall traffic volumes. In this work, the antenna height of BS utilized is 35 meters. And on the top of each vehicle, an isotropic antenna is installed at a height of 1.5 meters. And 1×2 antennas configuration (i.e., Rx diversity) is exploited for the direct C-V2X communication over sidelink. Also, each C-V2X Tx has a constant transmission power of 24 dBm. In addition, the transmission on PC5 is over 5.9 GHz central frequency with a bandwidth of 10 MHz.   
	
	\subsection{Traffic Model}\label{SCM}
	
	Traffic model specific for the safety-and-efficiency-related issues in the direct C-V2X communication includes both the event-driven and the periodical massages.
	
	\begin{itemize}
		\item Event-driven message: For this transmission, once a vehicle experiences certain events from the local environment, the event-driven messages will be delivered to all the vehicles in the proximity of the Tx. However, this type of message only needs to be generated and transmitted for one time, and thus a small volume of data traffic is expected.
		\item Periodic transmission: Compared to the event-driven traffic for the direct C-V2X communication, the periodic transmission refers to a continuously transmitting information including location, speed or roadway situation.
	\end{itemize}
	In this work, we utilize a periodic package transmission of 212 bytes with 10 Hz periodicity for each vehicle [11], since the traffic data volume of periodic transmission is higher than that of the event-driven messages. 
	
	\subsection{Modulation and coding schemes}
	An appropriate Modulation and Coding Scheme (MCS) is quite important for a point-to-multipoint communication and should be able to meet the system capacity requirement and provide a good robustness [3]. Thus, the MCS Spectral Efficiency (SE) is calculated as:
	\begin{equation}
	SE \geq S \times R \times N/BW \label{eq}
	\end{equation}
	Where $S$ represents the packets size and $R$ is the package  transmission period.   $BW$ is the allocated bandwidth and N is the number of UEs. In order to guarantee a robust transmission, an MCS with a lower spectral efficiency is required. So the network should apply the MCS which has the lowest spectral efficiency while fulfills the condition shown in Eq. (1). 
	
	\subsection{Key Factor Indicator (KPI)}
	In order to inspect the performance evaluation with the reliability by applying the three channel models, a value of 1\% is set as the threshold w.r.t. the (Block-Error-Ratio) BLER which means if the BLER value of each Rx is lower than the threshold value, then we assume the data packets transmitted by the Tx have been successfully received by the Rx. BLER and PRR are mutually complementary. So we use PRR values to represent the performances characteristics of the three channel models.
	
	\section{Channel Model}
	In general, radio propagation models can be fallen into two broad categories: large-scale propagation models and small-scale propagation models. 
	\begin{itemize}
	\item Large-scale propagation models: They are used to predict the average signal strength decays and combine overall effect of the pathloss and shadowing. 
	\item Small-scale fading models: They are used to characterize rapid fluctuations of the received signal strength over very short distance or very short time durations.
	\end{itemize}
	Fig.2 shows the combined effects of the large-scale propagation and small-scale propagation. It shows that the radio signal attenuation as a function of distance from the source may be conceived as the superposition of the pathloss, shadowing, and multipath effects [12]. 
	
	\begin{figure}[htbp]
		\centering
	    \includegraphics[width=\linewidth]{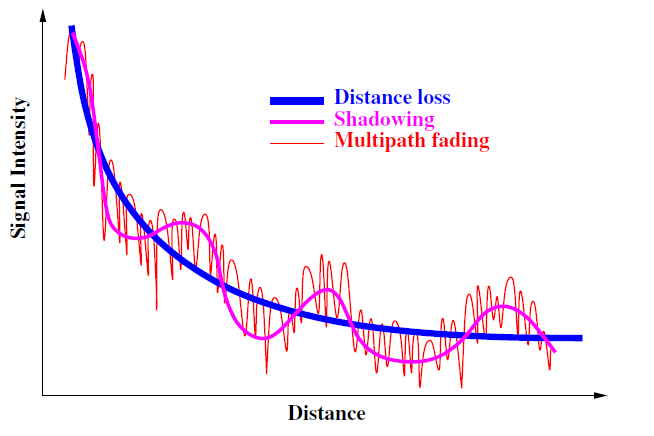}
        \caption{Combined Effects of Pathloss, Shadowing, and Multipath Fading [12].}
		\label{fig}
	\end{figure}
	There are many channel propagation models for cellular communication. Deriving channel models for C-V2X communication is still in initial stages. Channel propagation models for C-V2X communication are far more sophisticated. For the purpose of evaluation, we consider the following channel models for analyzing the performance of the direct C-V2X communication. In this work, only large-scale fading (e.g., pathloss and shadow fading) are exploited. 
	
	\subsection{Model 1: Two-Ray Interference Model}
	In this work, the Two-Ray interference model is used. Because this model is a realistic treatment of the pathloss which takes into account that the radio signal gets at least at the ground [13]. And the pathloss is calculated as:
	
	\begin{equation}
	PL = 10log10(\frac{4 \pi d}{\lambda})^2   \hspace{1.5cm}  if  \hspace{0.2cm}  d \leq d_c, \label{eq}
	\end{equation}
	\begin{equation}
	PL = 20log10(\frac{d^2}{h_{Tx}h_{Rx}})     \hspace{1cm}   if   \hspace{0.2cm} d > d_c, \label{eq}
	\end{equation}
	Where $h_{Tx}$ and $h_{Rx}$ are the antenna heights of the Tx and the Rx, and the d is the distance between the Tx and Rx.  $d_c$ is the cross distance for Two-Ray ground model and can be calculated as:
	\begin{equation}
	d_c = 4 \pi \frac{h_{Tx}h_{Rx}}{\lambda}.\label{eq}
	\end{equation}
	
	\subsection{Model 2: WINNER II Channel Models}
	The WINNER II channel models are propagation models for calculating the pathloss. WINNER II channel models can be used in both link-level and system-level simulations, as well as comparison of different channel models. That’s why we choose the WINNER II channel model for this system-level simulator. There are various propagation scenarios considered in WINNER II channel models, e.g., indoor office, urban macro-cell, or rural macro-cell [15]. In this work, since the simulation is based on a highway scenario, pathloss model for the rural macro-cell has been utilized in this case which is WINNER II D1 model. Both rural (Line of Sight) LOS and (None Line of Sight) NLOS states are both considered.

	\begin{multline}
	PL_{LOS} = 21.5log10(d) +20log10(\frac{f_c}{5.0}),   \hspace{0.5cm}   \sigma= 4,\\
	if \hspace{0.2cm} 10m < d < d_{BP}, \label{eq}
	\end{multline}
	\begin{multline}
	PL_{LOS} =  40log10(d) +10.5 - 18.5log10(h_{BS})-   \\
	18.5log10(h_{MS})+1.5log10(\frac{f_c}{5.0}),   \hspace{0.5cm}   \sigma= 6,\\
	if \hspace{0.2cm} d_{BP} < d < 10km, \label{eq}
	\end{multline}	
	\begin{multline}
	PL_{NLOS} =  25.1log10(d) +55.4 - 0.13log10(h_{BS}-    \\
	25)log10(\frac{d}{100})-0.9(h_{MS}-1.5)+21.3log10(\frac{f_c}{5.0}), \hspace{0.5cm}   \sigma= 8, \\
	if \hspace{0.2cm} 50m< d < 5km, \label{eq}
	\end{multline}
	Where $fc$ is the central frequency in 5.9 GHz and  $d$ is in meters. $d_{BP}$ is the breakpoint and computed as $d_{BP}$=4\(h_{BS}\)\(h_{MS}\)\(fc/c\) where $fc$ is in Hz. $h_{MS}$ and $h_{BS}$ are the antenna heights of the mobile station and BS. The shadowing models for the three cases are added to the direct transmission links which are log-normal random variables with 4 dB, 6 dB, and 8 dB standard deviations respectively.
	\subsection{Model 3: 3GPP Channel Model}
	In the latest 3GPP releases 15 [14], there is a new channel model modelled according to the LOS and NLOS states. Both LOS propagation and NLOS propagation modelled for the C-V2X sidelink are given as:
	\begin{equation}
	PL_{LOS} =32.4 + 20log10(d) +20log10(f_c),  \label{eq}
	\end{equation}
	\begin{equation}
	PL_{NLOS} =36.85 +18.9log10(f_c).  \label{eq}
	\end{equation}
	In Eq. (8) fc is the central frequency in 5.9 GHz and d is in meters.
	
	The shadow fading is also added to the transmission link which is a log-normal random variable with 3 dB standard deviation. In [13], an additional vehicle blockage loss has been introduced. So we also add this blockage loss in this work. 
	\section{Numerical Results}
	In this part, the performances of different C-V2X communication schemes are provided w.r.t PRRs. First of all, in Tab. I, we take the IVD of 10 meters with different resource allocation schemes applying different channel models into consideration. In order to inspect the system performance of the direct C-V2X communication, an efficient method is to apply different transmission bandwidths for the same channel model. As shown in Eq. (1), a larger system bandwidth can reduce the required spectral efficiency. Thus, an MCS value with a better robustness can be applied in this case. For example, when we increase the system bandwidth from 5 MHz to 10 MHz, the PRRs have been improved from 68.32 \% to 90.10 \%, from 39.83 \% to 54.85 \%, and from 76.73 \% to 97.86 \% corresponding to the Two-Ray interference model, WINNER II channel model, and 3GPP channel model. We can find out the different resource allocation schemes of the same channel model have a large impact on the system performance. Another efficient method is to apply different channel propagation models. It is worth noticing that the PRR values of the C-V2X communication using the 3GPP channel model are the largest among the three different channel models with same transmission bandwidth, which means the performance of the direct C-V2X communication system by applying the 3GPP channel model on a highway scenario can be improved a lot compared to Two-Ray interference model and WINNER II channel model. In Fig.3, the Cumulative Distribution Function (CDF) of the pathloss and shadowing of the three different channel propagation models is plotted. And the range of the distance between the Tx and an Rx is from 1 meter to 8000 meters. Since the fast movement of vehicles on a highway scenario requires a large C-V2X communication range, the estimated PRRs should be collected within a communication range of 1000 meters. It can be easily found that the 3GPP channel model has the lowest pathloss value among the three different channel models. That is why we can get the most reliable performance of the direct C-V2X communication with larger communication range requirement by applying the 3GPP channel model. For instance, if the transmission bandwidth is 10 MHz, the PRR is increased from 54.85\% using WINNER II channel model to 100\% using 3GPP channel model. So 3GPP channel model can be applied to C-V2X applications which require ultra-high reliability on a highway scenario. 
		
	In Tab. II, we increase the IVD from 10 meters to 15 meters which means fewer vehicles are deployed on the highway. With the decreased UEs, a lower system capacity is introduced here. So the MCS with better robustness can be applied. That is the reason why the PRRs in Tab. II are higher than those in Tab. I within the same channel model. Moreover, the performance of the direct C-V2X communication using the 3GPP channel Model is also the most practical and reliable in this case.
		
	Fig.3 illustrates the CDF of pathloss values with and without shadowing. We take the WINNER II channel model as an example. When the distance between the Rx and Tx is less than 1000 meters, the channel model with or without shadowing doesn’t affect the system performance. But if the distance increase up 6500 meters, the effect of shadowing fading on the system performance becomes more obvious than before. However, the effect of the shadowing on 3GPP channel model is less obvious than the WINNER II channel model with shadowing. 
	
	\begin{table}[htbp]
		\caption{Table I: System Performance of Different resource allocations Based on Different Channel Models without Shadowing(IVD=10Meters)}
		\begin{center}
			\begin{tabular}{|c|c|c|c|c|}
				\hline
				Index & BW & PRR Model 1& PRR Model 2 & PRR Model 3\\ 
				\hline
				1     & 5MHz  & 68.32\%   & 39.83\%  & 73.60 \% \\ 
				\hline
				2     & 6MHz  & 72.28\%   & 43.17\%  &  85.64\%   \\ 
				\hline
				3     & 8MHz  & 86.14\%    & 53.50\% &  100\%   \\
				\hline
				4     & 10MHz & 90.10\%    & 54.85\% &  100\%  \\
				\hline
			\end{tabular}
		\end{center}
	\end{table}
	\begin{table}[htbp]
		\caption{Table II: System Performance of Different resource allocations Based on Different Channel Models without Shadowing(IVD=15Meters)}
		\begin{center}
			\begin{tabular}{|c|c|c|c|c|}
				\hline
				Index & BW & PRR Model 1& PRR Model 2 & PRR Model 3\\ 
				\hline
				1     & 5MHz  & 86.14\%   & 55\%  & 90.35\% \\ 
				\hline
				2     & 6MHz  & 90.10\%   & 60.50\%  &  99.17\%   \\ 
				\hline
				3     & 8MHz  & 100\%    & 64.50\% &  100\%   \\
				\hline
				4     & 10MHz & 100\%    & 65.65\% &  100\%  \\
				\hline
			\end{tabular}
		\end{center}
	\end{table}
	
	  \begin{figure}[htbp]
		\centering
		\includegraphics[width=\linewidth]{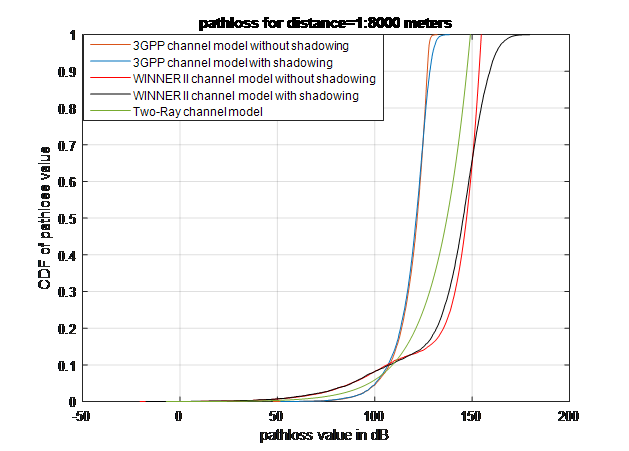}
		\caption{Pathloss of Different Channel Models.}
	    \label{fig}
	  \end{figure}
  
	\section{Conclusion}
	In this work, the performance of the direct C-V2X communication through sidelink has been analyzed under three different channel models. As some C-V2X communication applications require the ultra-high reliability, we analyze different channel models and try to find what kind of channel model can meet the requirements for the C-V2X applications on a highway scenario. Also, a detailed resource allocation scheme has been provided to analyze the system performance of the direct C-V2X communication. We also prove the shadowing for different channel models has different effects on the system performances. Further, according to the system performance characteristics of different channel models, we can apply the appropriate channel models for specific applications. So more researches on channel analyzing can be reduced. Last but not least, in order to evaluate the proposed technology, we have also implemented a system-level simulator with the simulation results. 
	\section*{Acknowledgment}
	A part of this work has been supported by Federal Ministry of Transport and Digital Infrastructure of the Federal Republic of Germany (BMVI) in the framework of the project ConVex with the funding number of 16AVF1019. The authors would like to acknowledge the contributions of their colleagues, although the authors alone are responsible for the content of the paper which does not necessarily represent the project.
	
\end{document}